\documentclass[%
 reprint,
superscriptaddress,
nofootinbib,
 amsmath,amssymb,
 aps,
prl,
]{revtex4-1}

\usepackage{siunitx}
\usepackage{graphicx}
\usepackage{dcolumn}
\usepackage{bm}
\usepackage{todonotes}
\usepackage{simplewick}
\usepackage{hyperref}
\usepackage{physics}
\usepackage{verbatim}
\usepackage{float}
\usepackage{array}
\usepackage{lipsum}
\usepackage{svg}
\hypersetup{
     colorlinks,
     linkcolor={blue!50!black},
     citecolor={blue!50!black},
     urlcolor={blue!80!black}
}
\graphicspath{{./figures/}}

\usepackage{bbm}
\usepackage{siunitx}
\usepackage{array}
\usepackage{xspace}
\usepackage{multirow}
\usepackage{tikz}
\usetikzlibrary{shapes.geometric,shapes.misc}
\makeatletter
\DeclareRobustCommand\onedot{\futurelet\@let@token\@onedot}
\def\@onedot{\ifx\@let@token.\else.\null\fi\xspace}

\makeatother

\newcommand{\like}{\mathcal{L}}

\newcolumntype{M}[1]{>{\centering\arraybackslash}m{#1}}

\begin{document}

\title{Ab initio short-range nuclear matrix elements for neutrinoless double-beta decay}%





\author{A.~Todd}%
\affiliation{TRIUMF, 4004 Wesbrook Mall, Vancouver, BC V6T 2A3, Canada}%
\affiliation{Department of Physics, McGill University, 3600 Rue University, Montr\'eal, QC H3A 2T8, Canada}%

\author{T.~Shickele}%
\affiliation{TRIUMF, 4004 Wesbrook Mall, Vancouver, BC V6T 2A3, Canada}%
\affiliation{Department of Physics \& Astronomy, University of British Columbia, Vancouver, BC V6T 1Z1, Canada}

\author{A.~Belley}%
\affiliation{Massachusetts Institute of Technology,
Cambridge, Massachusetts 02139, USA}%
\affiliation{TRIUMF, 4004 Wesbrook Mall, Vancouver, BC V6T 2A3, Canada}%
\affiliation{Department of Physics \& Astronomy, University of British Columbia, Vancouver, BC V6T 1Z1, Canada}

\author{L.~Jokiniemi}%
\affiliation{Technische Universit\"at Darmstadt, Department of Physics, D-64289 Darmstadt, Germany}%
\affiliation{ExtreMe Matter Institute EMMI, GSI Helmholtzzentrum f\"ur Schwerionenforschung GmbH, D-64291 Darmstadt, Germany}%
\affiliation{TRIUMF, 4004 Wesbrook Mall, Vancouver, BC V6T 2A3, Canada}%

\author{J.~D.~Holt}%
\affiliation{TRIUMF, 4004 Wesbrook Mall, Vancouver, BC V6T 2A3, Canada}%
\affiliation{Department of Physics, McGill University, 3600 Rue University, Montr\'eal, QC H3A 2T8, Canada}%

\begin{abstract}
We present converged ab initio calculations of short-range neutrinoless double-beta ($0\nu\beta\beta$) decay nuclear matrix elements for the key experimental isotopes $^{76}$Ge, $^{82}$Se, $^{130}$Te and $^{136}$Xe. 
Starting from different nuclear forces derived from chiral effective field theory, we apply the in-medium similarity renormalization group to obtain an effective valence-space Hamiltonian along with consistently transformed $0\nu\beta\beta$-decay operators.
We then obtain a range of values for the matrix elements that is consistent with, but generally smaller than, those from phenomenology. 
Finally, we combine our results with current limits from $0\nu\beta\beta$-decay searches to obtain constraints for the sterile-neutrino mixing-mass parameter space when considering the inclusion of a fourth sterile neutrino.

\end{abstract}

\maketitle

Neutrinoless double-beta ($0\nu\beta\beta$) decay is a hypothetical nuclear process in which two neutrons decay into two protons, emitting two electrons without associated antineutrinos~\cite{MasterFormula}. 
This decay is forbidden in the Standard Model (SM) of particle physics, as it violates lepton number but conserves baryon number, thus breaking $B$-$L$ symmetry~\cite{Deppisch2018}, a common feature among proposed beyond-Standard-Model (BSM) theories~\cite{PerezPavel}. 
In the SM neutrinos are massless left-handed Weyl fermions~\cite{Brdar2019}, but the existence of neutrino-flavour oscillation implies that neutrinos must have small masses~\cite{Simkovic2010, Bolton:2019}, leading to the question of whether neutrinos are Majorana fermions (i.e., their own antiparticles) or Dirac fermions (i.e., have distinct antiparticles)~\cite{Agostini2023}. 
Observation of $0\nu\beta\beta$ decay would both confirm the neutrino as Majorana and provide direct evidence of lepton-number violation (LNV). 

As $0\nu\beta\beta$-decay is an inherently BSM process, there are many proposed scenarios that could contribute to the physics of the decay. 
The standard mechanism, mediated by light-neutrino exchange, has been extensively studied, while possible contributions arising from  exotic LNV mechanisms remain less explored within nuclear theory~\cite{Horoi2017,Agostini2023}.
Many such mechanisms involve heavy degrees of freedom that, once integrated out at the quark level, induce LNV operators whose contributions at the nuclear scale are encoded in short-range nuclear matrix elements (NMEs) and low-energy constants (LECs)~\cite{MasterFormula}. 
Key scenarios involving heavy Majorana neutrinos, typically with right-handed (RH) chirality~\cite{Brdar2019} accompanied by a seesaw mechanism, are well motivated~\cite{deVries2024}. 
In addition to explaining the small masses of SM neutrinos, the type-I seesaw~\cite{MinkowskiPeter, Mohapatra1980} can simultaneously explain baryogenesis, the origin of the electroweak scale, and, provided one RH neutrino has a mass in the KeV range, potentially even dark matter~\cite{Brdar2019, Drewes2013}.

There is a major worldwide campaign aiming to discover $0\nu\beta\beta$-decay~\cite{GERDA2020,LEGEND2025,CUORE2024,Exo2019,KLZ2025}, with several next-generation ton-scale searches expected to be launched in the coming decades~\cite{nEXO2022,LEGEND2021,CUPID2024,SNO+2021}.
To date, constraints on BSM neutrino physics from experimental lifetime limits have largely focused on the standard mechanism.
 If nonstandard mechanisms contribute appreciably, however, $0\nu\beta\beta$-decay experiments can probe regions of parameter space inaccessible to many BSM searches; in particular, scenarios involving heavy Majorana neutrinos can extend sensitivity beyond the reach of high-energy colliders~\cite{Pascoli2019,Abdullahi2023,deLima2024}.

Our current capability to extract limits on light neutrino masses, the existence of new particles, and to differentiate between contributions from exotic LNV mechanisms~\cite{MasterFormula}, is severely hindered by our lack of knowledge concerning the NMEs, which govern the rate of the decay and can only be calculated from a nuclear theory perspective. 
NME calculations, for both standard and exotic LNV mechanisms, have long been an outstanding theoretical challenge~\cite{Agostini2023}, due to the need to consistently account for both nuclear and electroweak forces, within an accurate treatment of the nuclear many-body problem in heavy nuclei. 
Nuclear models have been employed to calculate NMEs in all key experimental isotopes, but there is notable disagreement between predictions from different methods~\cite{Menendez2018, Horoi2016, MENENDEZ2009139, Faessler2014, Fang2018, Hyvarinen2015, Barea2015, Rodriguez2010}, with no clear path to obtain systematic theoretical uncertainties~\cite{Yoshida2018, Horoi2022,Horoi2023,Castillo2025}. 

\begin{figure}[t!]
    \centering
    \includegraphics[width=\linewidth]{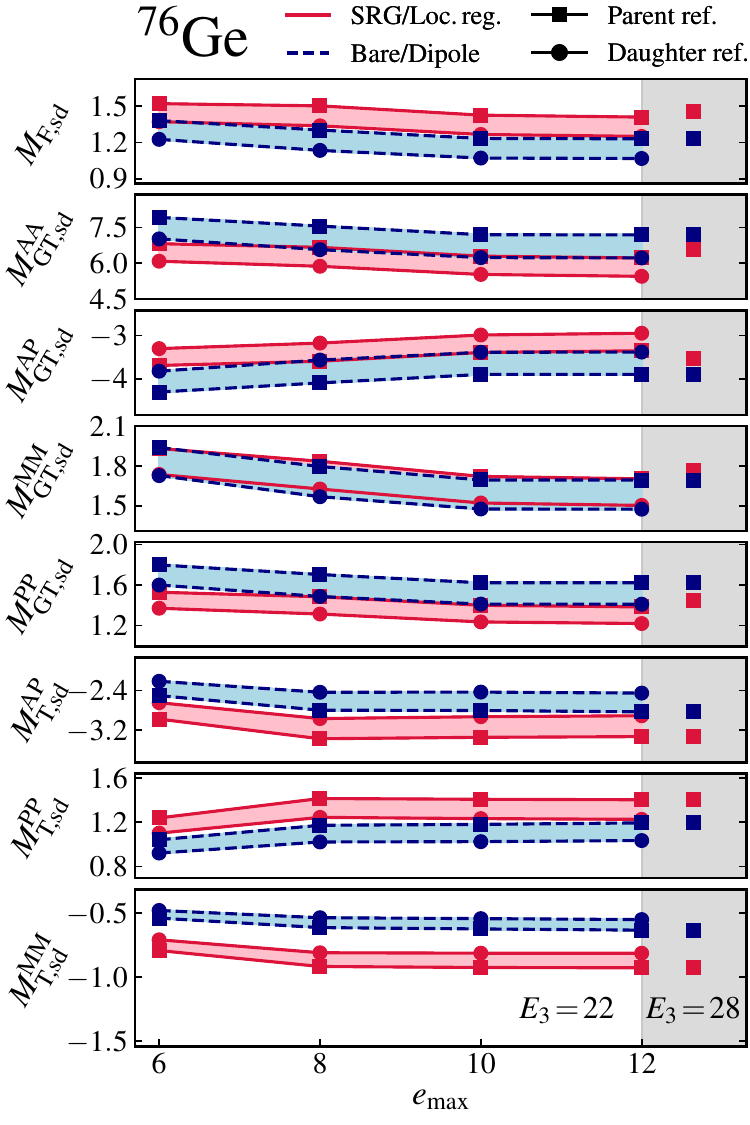}
    \caption{Convergence of short-range NMEs  in $^{76}$Ge as a function of $e_\text{max}$ and $E_\text{3max}$, with the N$^3$LO$_{\rm LNL}$ interaction, and two operator regularization schemes:~``SRG/Loc.~reg.'' and ``Bare/Dipole'' (see text). 
    In addition to NMEs computed using $E_\text{3max}\!=\!22$, we show NMEs at $e_\text{max}\!=\!12, E_\text{3max}\!=\!28$ in the grey regions. The bands represent a range due to the choice of ENO reference:~parent (square) or daughter (circle) nucleus. $M_{\rm F,sd}$ includes the factor $-(g_{\rm V}/g_{\rm A})^2$.}
    \label{fig:convergence}
\end{figure}

First-principles, or ab initio, theory instead approaches the nuclear many-body problem by taking all nucleons to interact via nuclear and electroweak forces based on chiral effective field theory (EFT)~\cite{Epelbaum2009,Machleidt2011}. 
These methods approximately solve the nuclear many-body problem within some nonperturbative, systematically improvable framework~\cite{Herg20Tour}, ultimately providing a clear prescription for rigorous uncertainty estimates.
Several ab initio methods show promising agreement for light-exchange NME benchmarks in light systems~\cite{Yao20Light},  $^{48}$Ca~\cite{Yao20Ca,Bell21Ge,Nova21Ca}, and the first uncertainty analysis for $^{76}$Ge~\cite{Bell2476Ge}.
In this Letter, we employ the Valence Space In-Medium Similarity Renormalization Group (VS-IMSRG)~\cite{Stroberg2019}, which has reached heavy candidates~\cite{Bell23TeXe}, to calculate short-range  NMEs needed for exotic LNV mechanisms mediating $0\nu\beta\beta$-decay for key experimental isotopes $^{76}$Ge, $^{82}$Se, $^{130}$Te, and $^{136}$Xe.
Then, assuming heavy-neutrino exchange dominates the total amplitude, we use current experimental likelihoods from $0\nu\beta\beta$-decay searches to obtain constraints in the sterile-neutrino mixing versus mass parameter space with the inclusion of a fourth sterile neutrino~\cite{Mohapatra1986,Okada1996}.

Like the nuclear forces, decay operators are formulated within chiral EFT, where high-scale LNV interactions are evolved to the nuclear scale. This yields a model-independent description of the decay rate, involving only a finite set of NMEs. Importantly, the nine long-range and six short-range NMEs describing light- and heavy-neutrino exchange are the same NMEs required for the leading-order contribution from any LNV source.

The short-range (sd) NMEs -- $M_{\rm F,sd}, M_{\rm GT,sd}^{\rm AA}, M_{\rm GT,sd}^{\rm AP}, $ $M_{\rm GT,sd}^{\rm PP}, M_{\rm T,sd}^{\rm AP}$ and $M_{\rm T,sd}^{\rm PP}$ -- correspond to Fermi (F), Gamow-Teller (GT), and Tensor (T) transitions, where superscripts indicate how operators transform under parity, either as axial (A), pseudoscalar (P), or magnetic (M) quantities, while the Fermi operator transforms as a vector (V). We also consider two additional matrix elements, $M_{\rm GT,sd}^{\rm MM}$ and $M_{\rm T,sd}^{\rm MM}$. Although they do not appear at leading order in the EFT power counting, we include them for comparison with previous phenomenological calculations~\cite{Menendez2018}. 

We define the total short-range NME as
\begin{align}
    \mathcal{M}^{0N} = -\left(\frac{g_{\rm V}}{g_{\rm A}}\right)^2 M_{\rm F,sd} + M_{\rm GT,sd} +  M_{\rm T,sd},
\end{align}
where $g_{\rm V}\!=\!1$, $g_{\rm A}\!=\!1.27$ are the unquenched vector and axial coupling constants, respectively, and each term is the sum of its AA/AP/PP/MM components, the expressions for which can be found in Appendix B. 
Although $\mathcal{M}^{0N}$ is not a meaningful operator in the EFT expansion, it provides a useful benchmark against nuclear models which use this combination \cite{Menendez2018}.

\begin{figure*}[t!]
  \centering  
  \includegraphics[width=\linewidth]{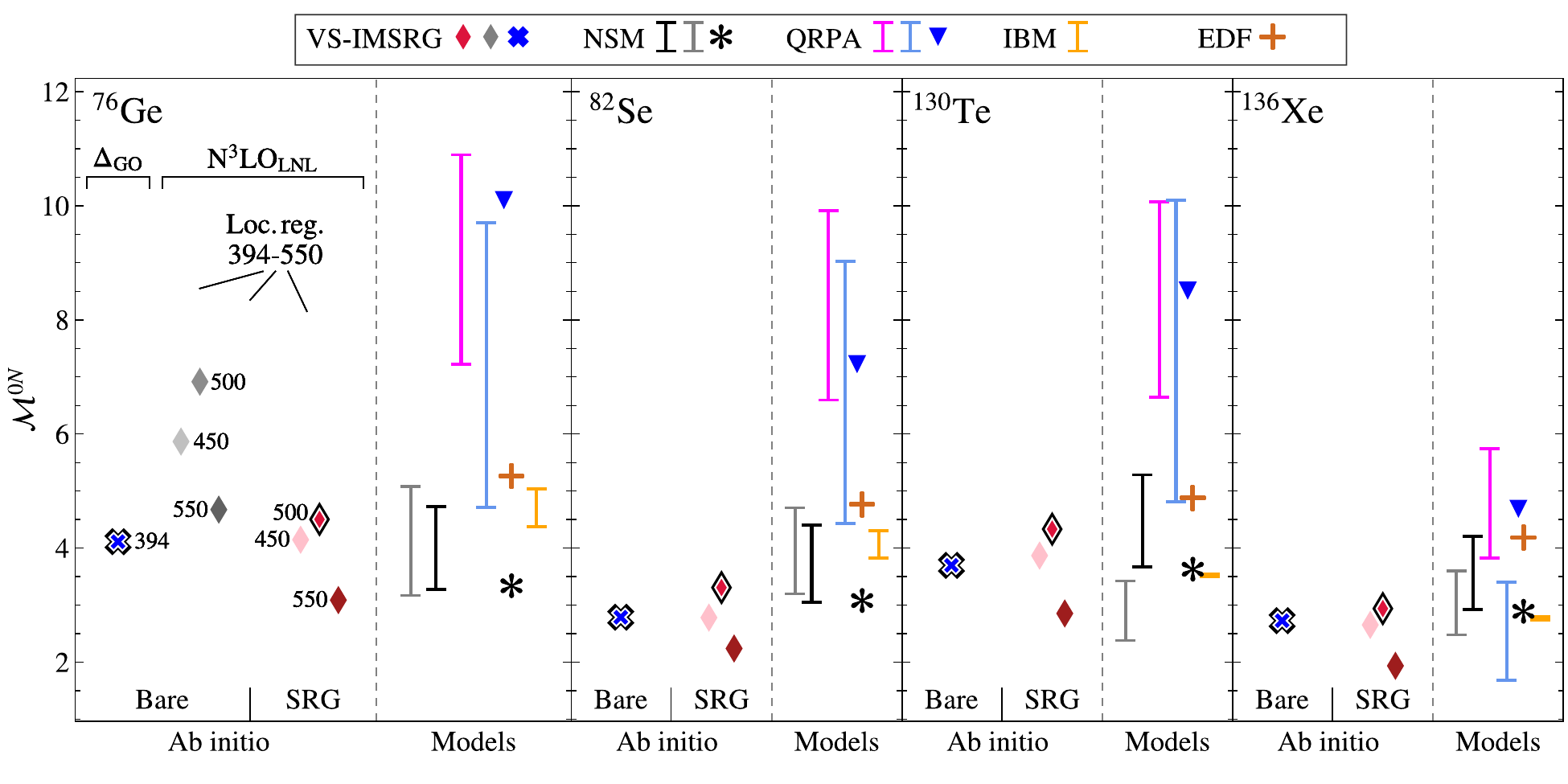}
  \caption{Comparison of the ab initio $\mathcal{M}^{0N}$ (N$^3$LO$_{\rm LNL}$ in diamonds and $\Delta_\text{GO}$ in crosses) with phenomenological results (``Models''). 
  Ab initio NMEs are split by bare or SRG-evolved operators, showing different regulator cutoffs and black outlines indicating operators are consistent with the interaction: N$^3$LO$_\text{LNL}$ with an SRG-evolved operator  with 500 MeV cutoff, and $\Delta_\text{GO}$ with a bare 394 MeV cutoff operator.
  Models include NMEs based on the NSM~\cite{Menendez2018,Horoi2016,MENENDEZ2009139}, QRPA~\cite{Faessler2014,Fang2018,Hyvarinen2015}, IBM~\cite{Deppisch2020}, and EDF~\cite{Song2017}.}
  \label{fig:comparison}
\end{figure*}

To explore the uncertainty arising from nuclear interactions, we work with two different sets of two- (NN) and three-nucleon (3N) forces based on chiral EFT. 
The N$^3$LO$_{\rm LNL}$ interaction includes NN and 3N forces consistently evolved using the similarity renormalization group (SRG) to a scale $\lambda\!=\!2.0$\,fm$^{-1}$, where the NN component uses a local regulator with cutoff 500\,MeV, and the 3N component uses a mixture of local and non-local regulators with cutoffs 650\,MeV and 500\,MeV, respectively~\cite{Leistenschneider2018,Soma20LNL}. 
In addition, the $\Delta_{\text{GO}}$ interaction explicitly includes $\Delta$ isobar degrees of freedom and uses a 394\,MeV cutoff for both NN and 3N forces at N$^2$LO and is not SRG-evolved~\cite{Jiang2020}. 
Both interactions share the same local regulator function in the NN sector (albeit with different cutoffs), denoted $f_\text{local}^{\text{NN}}(\textbf{q})$, whose form is given in Appendix B.
To be consistent with the SRG-evolved N$^3$LO$_{\rm LNL}$ interaction, we also co-evolve $0\nu\beta\beta$-decay operators, a step typically neglected for long-range NME calculations as this is expected to be negligible compared to other uncertainties~\cite{Parzuchowski2017}. 
Finally, we examine the effects of regulating $0\nu\beta\beta$-decay operators by either the same local regulator function $f_\text{local}^\text{NN}(\textbf{q})$ as in the NN sector, or the dipole parameterization of vector and axial form factors \cite{Simkovic2009}. 

We start from the harmonic-oscillator (HO) basis with frequency $\hbar\omega\!=\!15$\,MeV and restrict the single-particle space by $e\!=\!2n\!+\!l\!\leq\!e_\text{max}$ and 3N matrix elements by $e_1 + e_2 + e_3 \leq E_\text{3max}$, using the NuHamil code~\cite{Miyagi2023}.
We then free-space SRG evolve both the interaction and operators to a scale of $\lambda\!=\!2$\,fm$^{-1}$, where to quantify the effect of the SRG on the operators, we also calculate NMEs without this evolution. 
Transforming to the Hartree-Fock basis, we take 3N forces between valence nucleons into account via ensemble normal-ordering (ENO), which introduces a dependence on the chosen reference state~\cite{Stroberg2017}. 
Using the Magnus approach~\cite{Morris2015} of the VS-IMSRG~\cite{Hergert2016,Stroberg2017,Stroberg2019,Miyagi2020} in the IMSRG(2) approximation, implemented in the IMSRG++ code~\cite{Stroberg_IMSRG_2018}, we obtain quasi-unitary transformations that decouple a given core reference with associated valence-space Hamiltonian and transition operator from the full $A$-body space.
Finally, the Hamiltonian is diagonalized with the KSHELL code~\cite{SHIMIZU2019372} to obtain wave functions and operator expectation values for final NMEs.
For the $\Delta_\text{GO}$ interaction, we use $\hbar\omega\!=\!12$\,MeV and do not SRG-evolve the operators. 

In Fig.~\ref{fig:convergence}, we illustrate the convergence of short-range NMEs in $^{76}$Ge as a function of $e_\text{max}$ utilizing two operator renormalization schemes:~``SRG/Loc.\,reg.'' regulates and SRG evolves the operator consistently with the interaction, while ``Bare/Dipole'' employs the bare operator with the dipole parameterization. 
We note the ``Bare/Dipole'' scheme is inconsistent, but since it is commonly used in the literature, we include it for reference.
We see that NMEs for all short-range operators are well converged by $e_\text{max}\!=\!12$, where the model-space truncation error is smaller than the $<\!15\%$ uncertainty associated with the choice of reference state.
To examine convergence in terms of $E_{\text{3max}}$, we calculate the NMEs for $E_\text{3max} = 22 - 28$~\cite{Miya22Heavy}, finding minimal changes compared to the spread from the choice of reference state or other sources of uncertainty discussed below. 
For the remainder of this work, we use NMEs obtained with $e_\text{max}=12$, $E_{\text{3max}}\!=\!22$, given in Table~\ref{tab:h-NMEs} in Appendix A.
Finally, we see the difference between the two schemes can be up to $50\%$, indicating a sensitivity to the renormalization procedure, as expected for short-range operators. 
Analogous plots for $^{82}$Se, $^{130}$Te, and $^{136}$Xe are found in Supplemental Material~\cite{supp}.

In Fig.~\ref{fig:comparison} we show 
$\mathcal{M}^{0N}$ (for the parent reference) compared with phenomenological results from literature. 
For N$^{3}$LO$_{\rm LNL}$ we use SRG-evolved operators, regulated consistently by the local regulator $f_\text{local}^\text{NN}(\textbf{q})$ with $500$\,MeV cutoff, while for $\Delta_\text{GO}$, we use bare operators and the local regulator with a $394$\,MeV cutoff. 
These NMEs, obtained from operators consistently renormalized with the respective interaction, are highlighted with a black outline. 
We compare these results with NMEs in the literature from phenomenology:~the nuclear shell-model (NSM)~\cite{Menendez2018,Horoi2016,MENENDEZ2009139}, quasiparticle random-phase approximation (QRPA)~\cite{Faessler2014,Fang2018,Hyvarinen2015},  interacting boson model (IBM)~\cite{Deppisch2020}, and relativistic energy density functional (EDF) theory~\cite{Song2017}. 
We see from Fig.~\ref{fig:comparison} that consistently renormalized NMEs fall within the range of nuclear models, but with reduced spread.

To investigate the importance of renormalizing the operator consistently with the interaction, we evaluate NMEs using multiple regulator cutoffs for both the bare and SRG-evolved operators with N$^3$LO$_\text{LNL}$, which we also include in Fig.~\ref{fig:comparison} for $^{76}$Ge. 
In general the NMEs are quite sensitive to the cutoff, but those with SRG-evolved operators exhibit a reduced dependence compared to those with bare operators. 
This is expected, as the SRG suppresses high-momentum components, making the operators less sensitive to the choice of regulator cutoff, whereas the bare operators retain strong short-range components that are more directly affected by the cutoff. 
This trend is most evident when comparing the 450 and 500 MeV cutoffs, though a sizable gap persists between 500 and 550 MeV, even with the SRG-evolved transition operator. 
Importantly, the NMEs obtained with bare operators disagree with those obtained with SRG-evolved operators, demonstrating that a consistent SRG evolution of the short-range operators is essential when using SRG-evolved interactions.


\begin{figure}[t!]
    \includegraphics[width=\columnwidth]{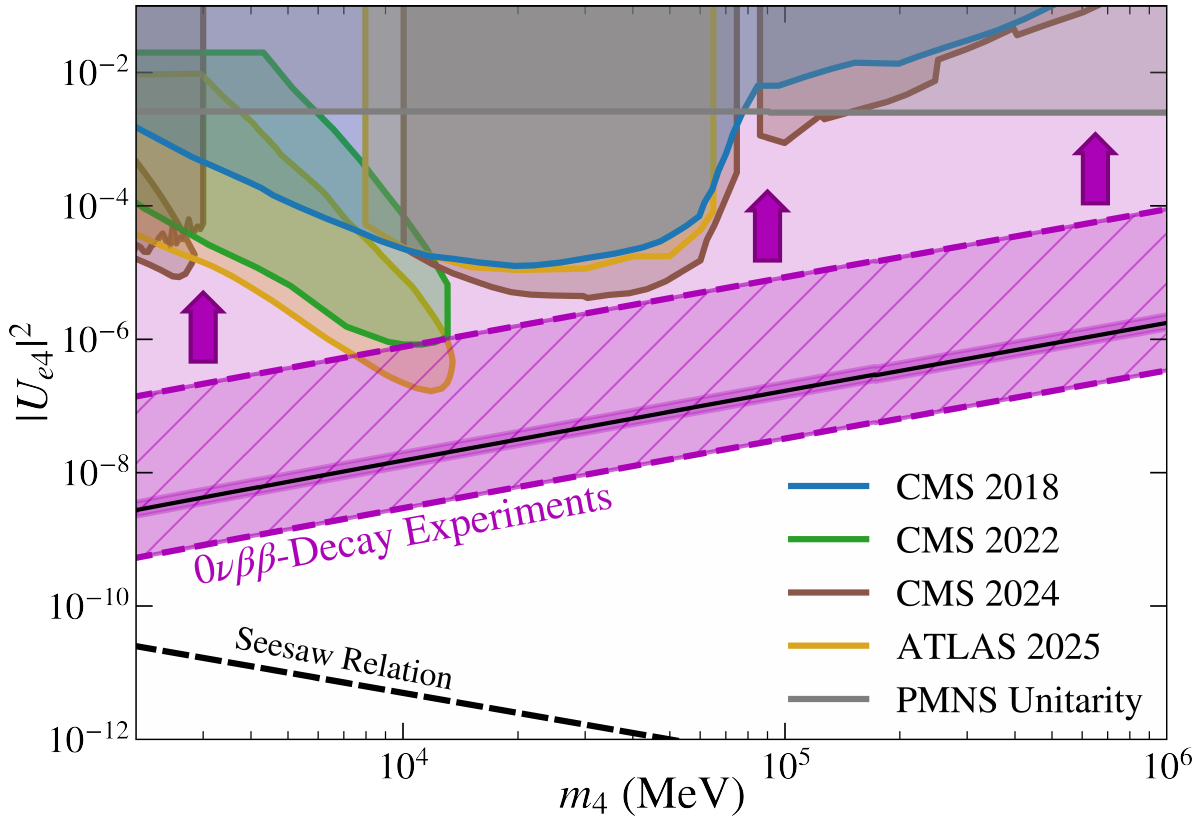}
    \caption{Constraints on $\big|U_{e4}\big|^2$ from a combination of current $0\nu\beta\beta$-decay experiments assuming a dominant contribution from heavy sterile neutrinos. Ab initio NME uncertainties are propagated into the darkest band surrounding the central limit (black line), with additional uncertainty from unknown LECs indicated by the hatched lines. Limits are compared against those from collider experiments~\cite{ATLAS2025_prompt,ATLAS2025_displaced,CMS2018,CMS2022,CMS2024_1,CMS2024_2,CMS2024_3} and PMNS unitarity~\cite{Blennow2023}. See text for further details.}
    \label{fig:heavy-neutrino-limits}
\end{figure}

We also note that, since the N$^3$LO$_\text{LNL}$ interaction uses a mixture of different cutoffs in the 3N sector, and these effects propagate to the two-body level through normal ordering, the most accurate cutoff for the transition operator likely varies from 500 MeV.
Although the NN and 3N forces are SRG-evolved to a common resolution scale of $\lambda\!=\!2\text{ fm}^{-1}$, which mitigates the regulator mismatch, the extent to which this mismatch affects external operators is not yet known. 
To assess this uncertainty, we include the range of values obtained by using 450--550 MeV cutoffs in the transition operators. 
Even in this conservative estimate, our NMEs still reduce the overall spread by a factor of 4 compared to nuclear models. 
For the $\Delta_\text{GO}$ interaction, there is no analogous cutoff uncertainty, as the same regulator cutoffs are used in both the NN and 3N sectors.
We exclude the EM(1.8/2.0) interaction~\cite{Hebeler2011,Simonis2017} from our analysis, since the NN force is SRG-evolved, while the 3N force is not, leaving no consistent framework to treat the transition operator. 
While our discussion so far has focused on interaction regulator dependence, at the EFT level, short-range operators also exhibit regulator dependence.  Integrating out high-energy degrees of freedom generates the short-range operators as well as LECs that run with the cutoff and absorb the regulator dependence of the NMEs (see the Supplemental Material~\cite{supp} for a more complete analysis). 


Finally, using our NMEs and experimental likelihoods from $0\nu\beta\beta$-decay searches, we compute $90\%$ credible interval combined Bayesian limits~\cite{Biller2021,Caldwell2017,Guzowski2015,Lisi2023} in the sterile--electron neutrino mixing--mass parameter space. 
For simplicity, we consider a fourth heavy sterile neutrino and assume heavy-neutrino exchange to be the dominant mechanism, such as can be realized in both the type-I seesaw mechanism due to cancellations in the light neutrino mass, and in the inverse and extended seesaw mechanisms from the decoupling of the light and heavy neutrino masses~\cite{Mitra2012,Lopez-Pavon2013}. 
The decay rate is then given by~\cite{Dekens2020,Dekens2024} 
\begin{align}
    \begin{split}&\left[T_{1/2}^{0\nu}\right]^{-1}= 4 g_{\rm A}^4 G_{01} V_{ud}^4 \eta(\mu,m_4)^2\big|U_{e4}\big|^4\frac{m_\pi^4}{m_e^2 m_4^2}\\
    &\ \ \ \ \ \ \ \ \ \times\bigg[\frac{5}{6}g_1^{\pi\pi}M_{\rm sd}^{\rm PP} + \frac{g_1^{\pi N}}{2}M_{\rm sd}^{\rm AP} + 2g_1^{NN}M_{\rm F,sd} \bigg]^2,\end{split}
    \label{eq:half-life}
\end{align}
where $g_{\rm A}$=1.27 is the axial-vector coupling constant~\cite{Gysbers2019}, $G_{01}$ the phase space factor~\cite{Stoica2019}, $V_{ud}$ is the up-down element of the Cabibbo-Kobayashi-Maskawa (CKM) matrix~\cite{Cabibbo1963,Kobayashi1973}, and $\eta(\mu,m_4)$ encodes the quantum chromodynamics (QCD) renormalization-group running~\cite{Dekens2024}. 
Here, $M_{\rm sd}^{\rm PP}$ and $M_{\rm sd}^{\rm AP}$ combine both GT and T components and $g_1^{\pi\pi}$, $g_1^{\pi N}$ and $g_1^{NN}$ are additional LECs.

In Fig.~\ref{fig:heavy-neutrino-limits} we employ a Bayesian prior uniform in decay rate $\Gamma^{0\nu}$, experimental likelihoods from LEGEND-200~\cite{LEGEND2025}, CUORE~\cite{CUORE2024}, EXO-200~\cite{Exo2019} and KamLAND-Zen~\cite{KLZ2025, ItaruShimizu} combined with computed NMEs~\cite{Biller2021,Shickele2026} to obtain global $0\nu\beta\beta$-decay constraints. 
Limits from ATLAS~\cite{ATLAS2025_displaced,ATLAS2025_prompt}, CMS~\cite{CMS2018,CMS2022,CMS2024_1,CMS2024_2,CMS2024_3}, PMNS unitarity~\cite{Blennow2023} constraints, and the seesaw relation, are also shown for comparison. 
Details on experimental likelihoods and limits are given in the Supplemental Material~\cite{supp}.
We take the range of results from N$^{3}$LO$_{\rm LNL}$ and $\Delta_{\rm GO}$, ENO reference, and operator cutoff (see Table~\ref{tab:h-NMEs}) as the NME uncertainty in Eq.~\eqref{eq:half-life}, illustrated in Fig.~\ref{fig:heavy-neutrino-limits} as the dark purple band surrounding the central limit (black line). 
For $g_1^{\pi\pi}$, we take the range between two recent Lattice QCD determinations $g_1^{\pi\pi}\!=\!0.36\pm\! 0.019$~\cite{Nicholson2018} and $g_1^{\pi\pi}\!=\!0.17 \!\pm\! 0.016$~\cite{Detmold2023}, and an estimate of $g_1^{NN}\!=\!(1\!+\!3g_A^2)/4$~\cite{Dekens2024}, with an uncertainty of $50\!-\!150\%$, is adopted for $g_1^{NN}$. 
Lastly, since $g_1^{\pi N}$ is unknown beyond its $\mathcal{O}(1)$ expectation, we take $g_1^{\pi N}\!\in\![\pm10^{-1/2},\pm10^{1/2}]$ to estimate uncertainty~\cite{Graf2022}. The uncertainties from LECs (mainly driven by $g_1^{\pi N}$) lead to the hatched purple band surrounding the central limit.
We see that while uncertainties in the final $0\nu\beta\beta$-decay constraints are dominated by the unknown LECs, ongoing searches competitively probe GeV-TeV scale sterile neutrinos. 
As the $0\nu\beta\beta$-decay mechanism is currently unknown, constraints may considerably weaken if heavy-neutrino exchange only contributes minimally.



In conclusion, we have calculated short-range $0\nu\beta\beta$-decay NMEs for $^{76}$Ge, $^{82}$Se, $^{130}$Te and $^{136}$Xe using the VS-IMSRG. 
We find convergence with respect to the single-particle space as well as a reduced spread across different chiral interactions compared to phenomenological estimates---even when accounting for the uncertainty associated with the regulator cutoff in the transition operator. 
We then derive global constraints on the sterile neutrino mixing $\big|U_{e4}\big|^2$ assuming $0\nu\beta\beta$-decay is mediated by heavy-neutrino exchange. Constraints are competitive with traditional collider and PMNS unitarity bounds, illustrating the potential for $0\nu\beta\beta$-decay to complement searches for heavy neutrinos.
The path towards final short-range NMEs requires addressing several ingredients, such as an improved treatment of many-body correlations with some approximation of IMSRG(3)~\cite{Hein21IMSRG3,He243f2}, inclusion of sub-leading $0\nu\beta\beta$-decay operators, and a deeper understanding of regulator-related uncertainties.
For a final rigorous error quantification, machine-learning emulators~\cite{Belley2026,Munoz26FRAME} provide a powerful framework to propagate uncertainties from interactions, operators, and many-body truncations to final observables. 
We also note that while we have so far only considered light- and heavy-neutrino-exchange mechanisms of $0\nu\beta\beta$-decay, analysis of more exotic scenarios, such as light-sterile-neutrino exchange \cite{Dekens2024}, are in progress. 
Ultimately, once these developments are achieved, short-range $0\nu\beta\beta$-decay NMEs will become quantitatively reliable inputs for interpreting a potential experimental discovery.

\begin{acknowledgments}

We thank V.~Cirigliano, J.~Detwiler, M.~Drissi, J.~Men\'{e}ndez, T.~Miyagi, S.R.~Stroberg, W.~Dekens, and D.~Tuckler for insightful discussions. 
The VS-IMSRG code employed in this work utilizes the Armadillo C$++$ library~\cite{Sanderson2016,Sanderson2018}. 
TRIUMF receives funding via a contribution agreement through the National Research Council of Canada. 
This work was supported by the Natural Sciences and Engineering Research Council of Canada (NSERC) under grant SAPIN-2024-0003, the Arthur B.~McDonald Canadian Astroparticle Physics Research Institute and the Canadian Institute for Nuclear Physics. 
A.B.~acknowledges support of NSERC [PDF-587464-2024]. 
L.J.~was supported by the LOEWE Top Professorship LOEWE/4a/519/05.00.002(0014)98 by the State of Hesse. 
Computations were performed with an allocation of computing resources on Cedar at WestGrid and the Digital Research Alliance of Canada. 

\end{acknowledgments}

\bibliography{library}

\appendix

\onecolumngrid

\section{End Matter}

\twocolumngrid
 
\textit{Appendix A: Table of NMEs}---we show in Table \ref{tab:h-NMEs} the short-range NMEs computed in this work, for the $\Delta_{\rm GO}$ and N$^3$LO$_{\rm LNL}$ chiral interactions, in the isotopes $^{76}$Ge, $^{82}$Se, $^{130}$Te and $^{136}$Xe.

\begin{table*}[]
    \centering
    \setlength{\tabcolsep}{5.5pt}
    
    \begin{tabular}{cS[table-format=1.3(2)]S[table-format=1.3(2)]S[table-format=1.3(2)]S[table-format=1.3(2)]S[table-format=1.3(2)]S[table-format=1.3(2)]S[table-format=1.3(2)]S[table-format=1.3(2)]S[table-format=1.3(2)]S[table-format=1.3(2)]S[table-format=1.3(2)]}
    \hline\hline
         & \multicolumn{2}{c}{$^{76}$Ge}& &\multicolumn{2}{c}{$^{82}$Se} & &\multicolumn{2}{c}{$^{130}$Te} & &\multicolumn{2}{c}{$^{136}$Xe} \rule[-1ex]{0pt}{4ex}\\
         & \multicolumn{1}{c}{$\Delta_\text{GO}$}&\multicolumn{1}{c}{N$^3$LO$_\text{LNL}$}& &  \multicolumn{1}{c}{$\Delta_\text{GO}$}&\multicolumn{1}{c}{N$^3$LO$_\text{LNL}$}& &\multicolumn{1}{c}{$\Delta_\text{GO}$}&\multicolumn{1}{c}{N$^3$LO$_\text{LNL}$} & &\multicolumn{1}{c}{$\Delta_\text{GO}$}&\multicolumn{1}{c}{N$^3$LO$_\text{LNL}$} \\
         \cline{2-3} \cline{5-6}\cline{8-9} \cline{11-12}
         $M_{\rm F,sd}$ & 0.93(0.04) & 1.1(0.3) & & 0.682(0.009) & 1.0(0.3) & & 0.89(0.03) & 1.3(0.3) & & 0.68(0.02) & 0.9(0.2) \rule[-1ex]{0pt}{4ex}\\
$M_{\rm GT,sd}^{\rm AA}$ & 5.5(0.3) & 5.4(0.8) & & 3.91(0.05) & 4.6(1.1) & & 5.3(0.2) & 6.1(0.9) & & 4.0(0.1) & 4.3(0.7)\rule[-1ex]{0pt}{4ex}\\
$M_{\rm GT,sd}^{\rm AP}$ &-2.8(0.1) & -2.9(0.4) & & -1.99(0.03) & -2.5(0.6) & & -2.7(0.1) & -3.4(0.5) & & -2.05(0.06) & -2.4(0.4)\rule[-1ex]{0pt}{4ex}\\
$M_{\rm GT,sd}^{\rm PP}$ &  1.08(0.05) & 1.2(0.2) & & 0.78(0.01) & 1.0(0.2) & & 1.06(0.04) & 1.4(0.2) & & 0.81(0.02) & 1.0(0.2) \rule[-1ex]{0pt}{4ex}\\
$M_{\rm GT,sd}^{\rm MM}$ & 0.93(0.05) & 1.3(0.4) & & 0.675(0.008) & 1.2(0.4) & & 0.92(0.03) & 1.6(0.5) & & 0.70(0.02) & 1.2(0.3) \rule[-1.5ex]{0pt}{4.5ex}\\ 
$M_{\rm T,sd}^{\rm AP}$ & -2.2(0.1) & -2.9(0.4) & & -1.7111(0.0002) & -2.6(0.7) & & -2.4(0.1) & -4.1(0.6) & & -1.92(0.06) & -3.1(0.5) \rule[-1ex]{0pt}{4ex}\\
$M_{\rm T,sd}^{\rm PP}$ & 0.90(0.05) & 1.2(0.2) & & 0.6875(0.0001) & 1.1(0.3) & & 0.97(0.04) & 1.7(0.3) & & 0.77(0.02) & 1.3(0.2) \rule[-1.5ex]{0pt}{4.5ex}\\ 
$M_{\rm T,sd}^{\rm MM}$ & -0.40(0.02) & -0.7(0.2) & & -0.3060(0.0002) & -0.7(0.3) & & -0.43(0.02) & -1.1(0.3) & & -0.34(0.01) & -0.8(0.2) \rule[-1.5ex]{0pt}{4.5ex}\\ \hline 
$\mathcal{M}^{0N}$ &3.9(0.2) & 3.8(0.7) & & 2.73(0.05) & 3.1(0.8) & & 3.6(0.1) & 3.6(0.7) & & 2.66(0.07) & 2.4(0.5) \rule[-1.35ex]{0pt}{4.5ex}\\
         \hline\hline
    \end{tabular}
    \caption{Fully-converged $e_\text{max}=12$, $E_\text{3max}=22$ short-range $0\nu\beta\beta$-decay NMEs for the isotopes $^{76}$Ge, $^{82}$Se, $^{130}$Te and $^{136}$Xe, including the total matrix element $\mathcal{M}^{0N}$. For each isotope we show NMEs obtained using the $\Delta_\text{GO}$ and N$^3$LO$_\text{LNL}$ chiral interactions. The uncertainty represents the uncertainty due to the choice of ENO reference for the $\Delta_\text{GO}$ interaction, and for the N$^3$LO$_{\rm LNL}$ interaction this uncertainty also factors in the range due to using different operator cutoffs (450-550 MeV) when using the parent nucleus as reference. The matrix element $M_{\rm F,sd}$ includes the factor of $-({g_{\rm V}}/{g_{\rm A}})^2$.}
    \label{tab:h-NMEs}
\end{table*}

\textit{Appendix B: Short-range operators}---the NMEs computed in this work are given by
\begin{align}
      M_{\alpha,\rm sd} =  \bigg\langle0_f^+ \bigg\| \sum_{m,n} O_\alpha^{mn}(\textbf{q})\bigg\|0_i^+\bigg\rangle
\end{align}
with the short-range $0\nu\beta\beta$-decay operators in momentum space,
\begin{align}
    O^{mn}_{\alpha} (\textbf{q})= \frac{R}{2\pi m_{\pi}^2} h_{\alpha}(\textbf{q}^2) S_\alpha (\textbf{q}) \tau^+_m \tau^+_n,
\end{align}
where $\alpha \in \{\rm F, GT, T\}$, $R=1.2A^{1/3}$ is the nuclear radius, $q$ is the momentum transfer between the nucleons $m$ and $n$, $\tau^+$ is the isospin-raising operator, and $m_\pi =  138.039$ MeV is the average pion mass. The factor $m_{\pi}^2$ was introduced to make the NMEs dimensionless and $\mathcal{O}(1)$ to correspond to the $\chi$EFT power counting~\cite{MasterFormula}. Note that this differs from the convention used in the literature~\cite{Menendez2018,Horoi2016,Faessler2014,Fang2018,Hyvarinen2015,Barea2015,Rodriguez2010}, by a factor of $m_{\rm p}m_{\rm e}/m_{\pi}^2$. The spin-spatial part of the operators, $S_\alpha(\textbf{q})$, is given by 
\begin{align}
    S_{\rm F} (\textbf{q}) &= \mathbbm{1},\\
    S_{\rm GT} (\textbf{q}) &= \boldsymbol{\sigma}_{m} \cdot \boldsymbol{\sigma}_{n},\\
    S_{\rm T}(\textbf{q}) &= -3 \frac{(\boldsymbol{\sigma}_m\cdot \textbf{q})(\boldsymbol{\sigma}_n\cdot \textbf{q})}{q^2} + \boldsymbol{\sigma}_{m} \cdot \boldsymbol{\sigma}_{n},
\end{align}
where $\boldsymbol{\sigma}$ are the Pauli matrices. The neutrino potentials $h_\alpha(\textbf{q}^2)$ are defined as 
\begin{align}
     h_{\rm F}(\textbf{q}^2) &= g_{\rm V}(\textbf{q}^2),\\
     h_{\rm GT}(\textbf{q}^2)&=h_{\rm GT}^{\rm AA}(\textbf{q}^2)+h_{\rm GT}^{\rm AP}(\textbf{q}^2)+h_{\rm GT}^{\rm PP}(\textbf{q}^2)+h_{\rm GT}^{\rm MM}(\textbf{q}^2),\\
     h_{\rm T}(\textbf{q}^2)&=h_{\rm T}^{\rm AP}(\textbf{q}^2)+h_{\rm T}^{\rm PP}(\textbf{q}^2)+h_{\rm T}^{\rm MM}(\textbf{q}^2),
\end{align}
where the different components are defined as
\begin{align}
    h_{\rm GT}^{\rm AA}(\textbf{q}^2) &= h_{\rm T}^{\rm AA}(\textbf{q}^2) = \frac{g_{\rm A}^2(\textbf{q}^2)}{g_{\rm A}^2},\label{np1}\\
    h_{\rm GT}^{\rm AP}(\textbf{q}^2) &= -h_{\rm T}^{\rm AP}(\textbf{q}^2)=\frac{g_{\rm P}(\textbf{q}^2)}{g_{\rm A}^2} g_{\rm A}(\textbf{q}^2)\frac{\textbf{q}^2}{3m_{\rm nucl}},\label{np2}\\
    h_{\rm GT}^{\rm PP}(\textbf{q}^2) &= -h_{\rm T}^{\rm PP}(\textbf{q}^2)=\frac{g_{\rm P}^2(\textbf{q}^2)}{g_{\rm A}^2}\frac{\textbf{q}^4}{12m_{\rm nucl}^2},\label{np3}\\
    h_{\rm GT}^{\rm MM}(\textbf{q}^2) &= 2h_{\rm T}^{\rm MM}(\textbf{q}^2) = g_{\rm M}^2(\textbf{q}^2) \frac{\textbf{q}^2}{6g_{\rm A}^2 m_{\rm nucl}^2}\label{np4} ,
\end{align}
with the average nucleon mass $m_{\rm nucl} = 938.919$ MeV and the vector, axial-vector, induced pseudo-scalar and weak-magnetism coupling constants $g_{\rm V},\  g_{\rm A}, \ g_{\rm P}$ and $g_{\rm M}$. We assume $g_{\rm P}(\textbf{q}^2)$ and $g_{\rm M}(\textbf{q}^2)$ are given by 
\begin{align}
    g_{\rm P}(\textbf{q}^2) &= -\frac{2m_{\rm nucl} g_{\rm A}(\textbf{q}^2)}{\textbf{q}^2 + m_\pi^2},\\
    g_{\rm M}(\textbf{q}^2) &= (1+\kappa_1) g_{\rm V}(\textbf{q}^2)
\end{align}
where $\kappa_1 = 3.7$ is the anomalous
isovector nucleon magnetic moment. The potentials $h_{\rm GT}^{\rm MM}(\textbf{q}^2)$ and $h_{\rm T}^{\rm MM}(\textbf{q}^2)$ are uniquely suppressed by $\textbf{q}^2/m_{\rm nucl}^2$ relative to other potentials, but this suppression is reduced by the factor $(1+\kappa_1)^2 \sim 22$. This enhancement is generic to both long- and short-range operators, however there is additional enhancement unique to short-range operators coming from  the suppression factor $\textbf{q}^2/m_{\rm nucl}^2$ being larger for short-range operators due to larger momentum transfers.

When we use the dipole parameterization of the vector and axial form factors, we have~\cite{MasterFormula}
\begin{align}
      g_{\rm V}(\textbf{q}^2) &= g_{\rm V}\left(1+\frac{\textbf{q}^2}{\Lambda^2_{\rm V} }\right)^{-2}, \\ g_{\rm A}(\textbf{q}^2) &= g_{\rm A} \left(1+\frac{\textbf{q}^2}{\Lambda^2_{\rm A}}\right)^{-2}
\end{align}
with $\Lambda_{\rm V} = 850$ MeV and $\Lambda_{\rm A} = 1086$ MeV. 

In an EFT framework, dipole form factors for the vector, axial and magnetic currents are not leading order and resum only a subset of higher-order contributions, thus providing no systematic improvement when applied to leading-order operators \cite{Cirigliano2018a}.  
While the $\textbf{q}^2$-dependence arising from the pion pole in the pseudoscalar form factor is genuinely leading order, promoting $g_\text{V}\!\to\! g_\text{V}(\textbf{q}^2)$ or $g_\text{A}\!\to\!g_\text{A}(\textbf{q}^2)$ is not justified. 
Therefore, in this work we retain only leading-order constants in the vector and axial currents and regulate the operator consistently with the interaction. In other words, we have
\begin{align}
    g_{\rm V}(\textbf{q}^2) &=g_\text{V} ,\label{eqs:gv}\\
    g_{\rm A}(\textbf{q}^2) &= g_{\rm A} \label{eqs:ga}
\end{align}
where $g_\text{V}=1,\  g_\text{A} = 1.27$ and we include a local regulator function $f_\text{local}^\text{NN}(\textbf{q})$ multiplying each neutrino potential (Eqs. (\ref{np1}-\ref{np4})). Specifically, we multiply by
\begin{align}
    f_\text{local}^\text{NN} (\textbf{q}) &= \exp \left[-\left(\frac{\textbf{q}}{\Lambda}\right)^{2n}\right]
\end{align}
with $\Lambda = 500 $ MeV and $n = 3$ to be consistent with the N$^3$LO$_\text{LNL}$ interaction and $\Lambda = 394$ MeV and $n=4$ to be consistent with the $\Delta_\text{GO}$ interaction. 
Using Eqs.~(\ref{eqs:gv}-\ref{eqs:ga}), we retain the leading-order pseudo-scalar and magnetic currents, $g_\text{P}(\textbf{q}^2)$ and $g_\text{M}(\textbf{q}^2)$, consistent in chiral perturbation theory.
\newpage

\clearpage

\setcounter{page}{1}
\begin{widetext}
\section*{Supplemental Material: Ab initio short-range nuclear matrix elements for neutrinoless double-beta decay}

\end{widetext}

\setcounter{figure}{0}
\renewcommand{\thefigure}{S\arabic{figure}}
\renewcommand{\theHfigure}{S\arabic{figure}} 

\setcounter{table}{0}
\renewcommand{\thetable}{S\arabic{table}}
\renewcommand{\theHtable}{S\arabic{table}}

\setcounter{equation}{0}            
\renewcommand{\theequation}{S\arabic{equation}}  
\renewcommand{\theHequation}{S\arabic{equation}} 

\section{Convergence plots for $^{82}$Se, $^{130}$Te and $^{136}$Xe} 
\begin{figure}[t!]
    \centering
  \includegraphics[width=\linewidth]{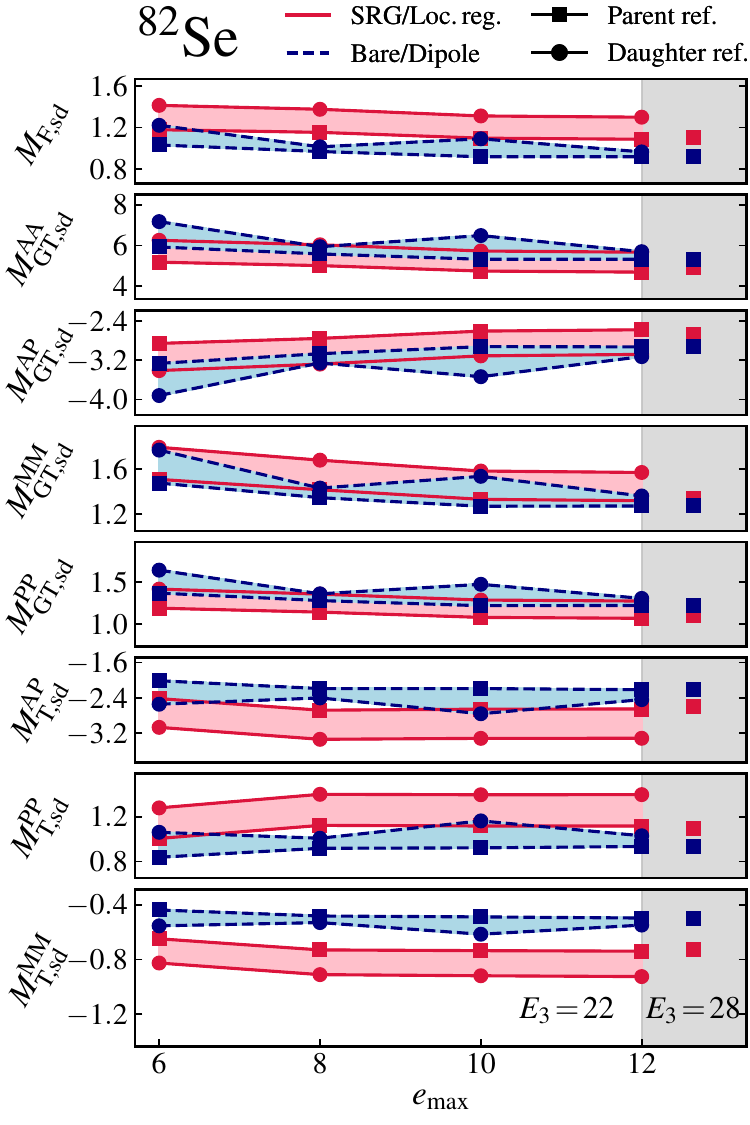}
    \caption{Same as Fig. \ref{fig:convergence} but for the isotope $^{82}$Se.}
    \label{fig:Se82}
\end{figure}

\begin{figure}[t!]
    \centering
  \includegraphics[width=\linewidth]{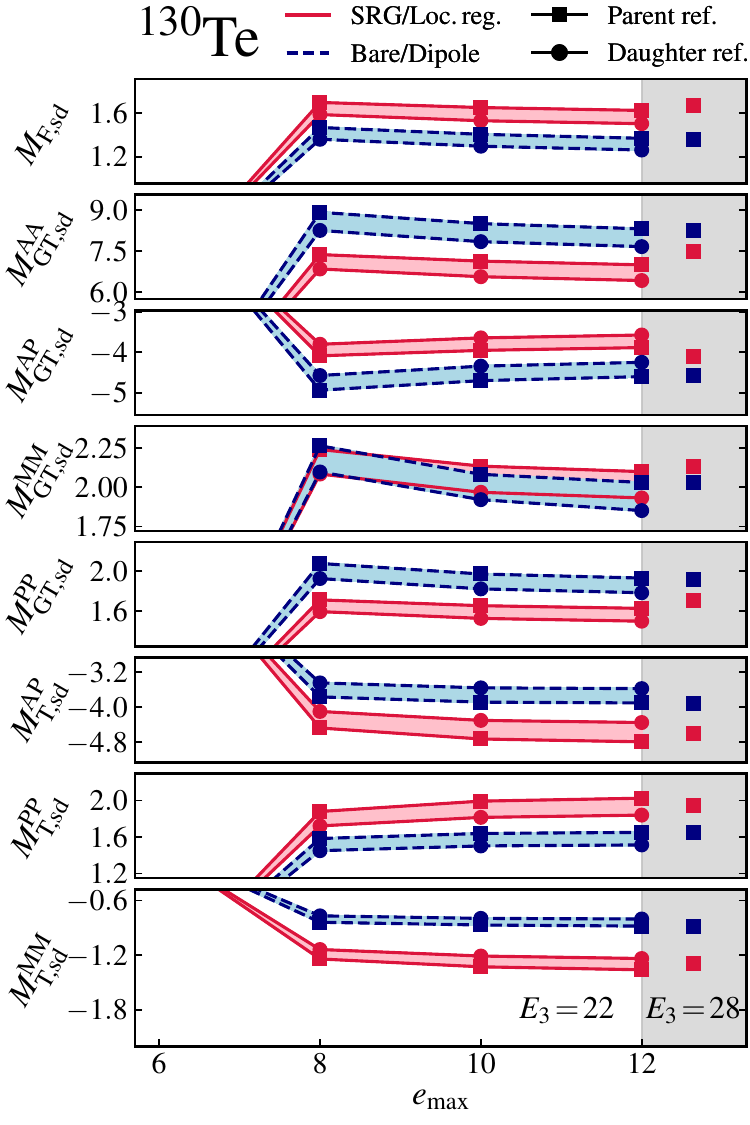}
    \caption{Same as Fig. \ref{fig:convergence} but for the isotope $^{130}$Te.}
    \label{fig:Te130}
\end{figure}

\begin{figure}[t!]
    \centering
  \includegraphics[width=\linewidth]{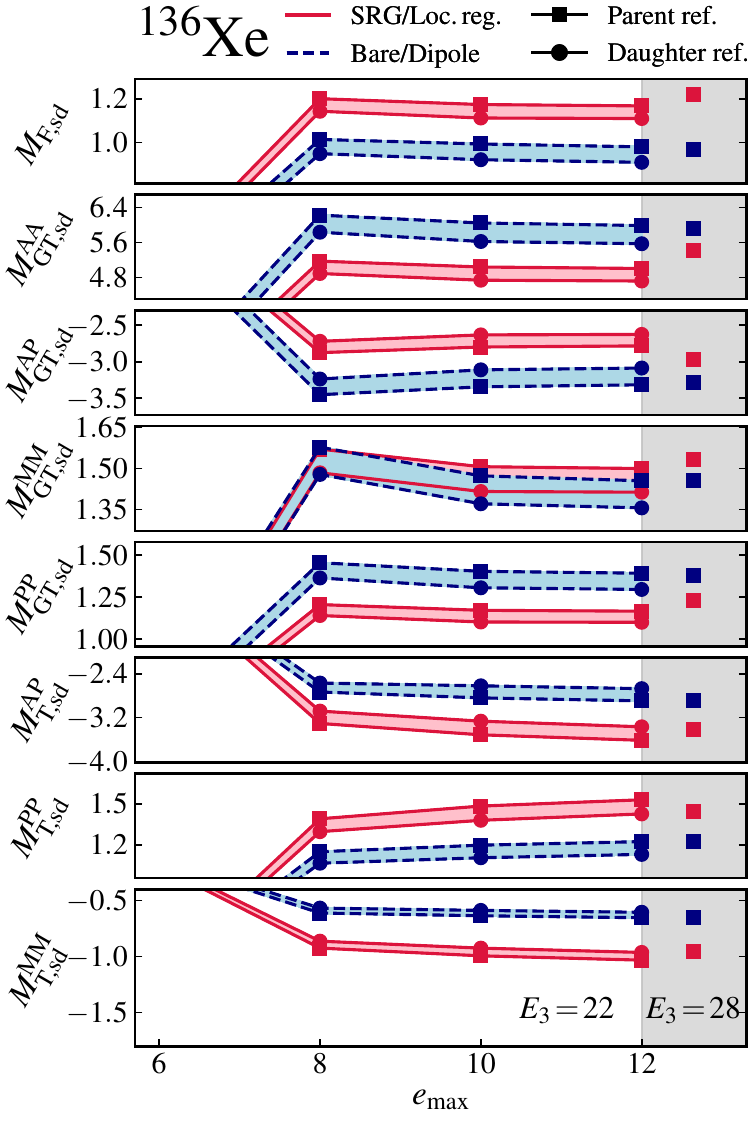}
    \caption{Same as Fig. \ref{fig:convergence} but for the isotope $^{136}$Xe.}
    \label{fig:Xe136}
\end{figure}

In this section, we provide plots of the short-range NMEs as a function of $e_\text{max}$ for $^{82}$Se, $^{130}$Te and $^{136}$Xe in Figs.~\ref{fig:Se82}, \ref{fig:Te130} and \ref{fig:Xe136}, and summarize the valence spaces used. 
For $^{76}$Ge and $^{82}$Se we take a $^{56}$Ni core with valence orbitals $f_{5/2}$, $p_{1/2}$, $p_{3/2}$ and $g_{9/2}$ for both protons and neutrons. For $^{130}$Te and $^{136}$Xe, we take a $^{100}$Sn core with valence orbitals $g_{7/2}$, $d_{3/2}$, $d_{5/2}$, $s_{1/2}$, $h_{11/2}$ for both protons and neutrons.
We see similar convergence for $^{82}$Se, $^{130}$Te and $^{136}$Xe as for $^{76}$Ge in Fig.~\ref{fig:convergence}. We note that the reference-state dependence in $^{82}$Se is different from the others in that, at $e_\text{max}=10$, the value of the NMEs changes substantially when using the daughter nucleus as the reference state. This behaviour is attributed to reodering of the orbitals from the mean field. Due to the deformation of $^{82}$Se, the NME is very dependent on the reference state and a small reordering can lead to large variation. Rather than fixing the ordering to remove this artifact, we present it here to highlight the larger many-body uncertainties in the case of deformed nuclei.

We also note that, in all isotopes, NMEs involving magnetic currents, $M_{\rm GT,sd}^{\rm MM}$ and $M_{\rm T,sd}^{\rm MM}$ are of comparable magnitude to other components, meaning their contribution is non-negligible. 
This confirms that, despite formal chiral suppression, the large isovector magnetic moment of the nucleon enhances these terms to the level of other contributions.

\begin{figure}
    \centering
    \includegraphics[width=\linewidth]{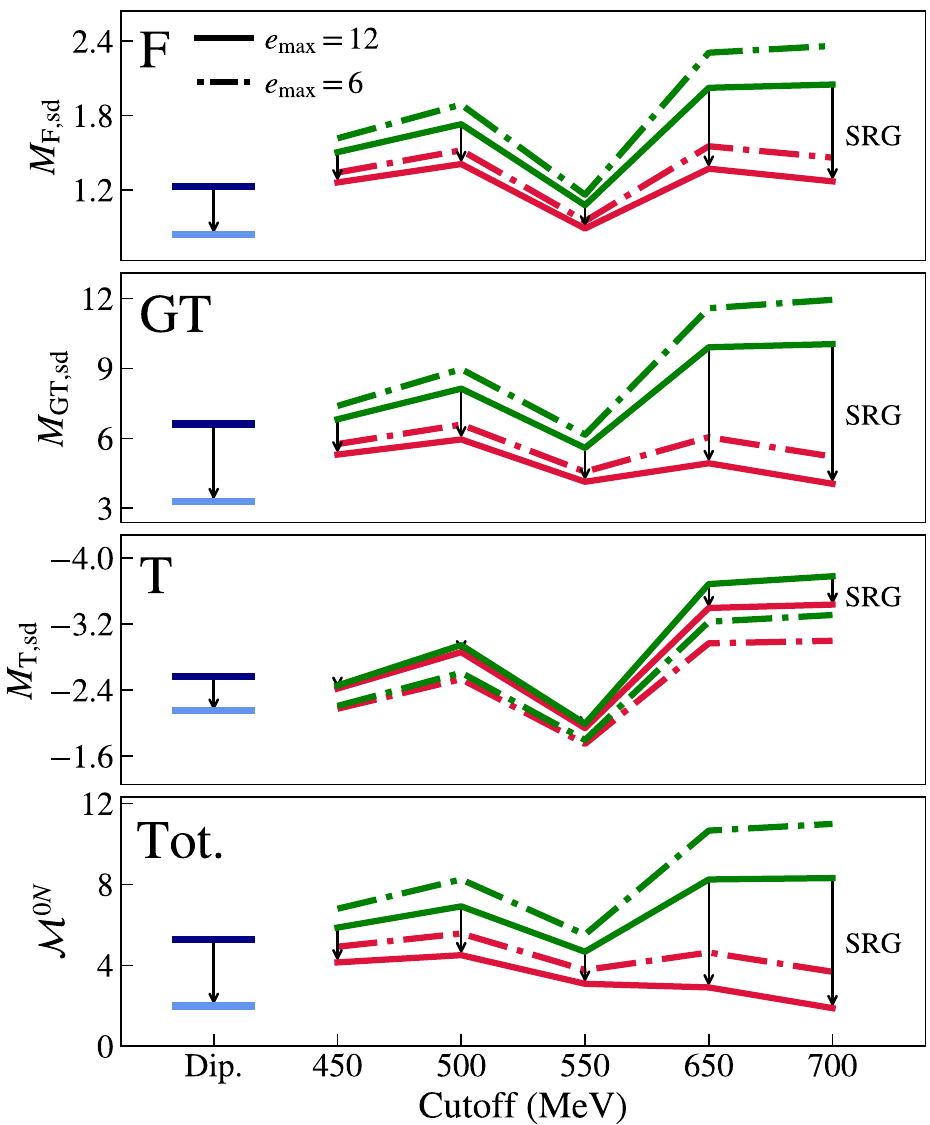}
    \caption{Plots of the matrix elements $M_{\rm F,sd}$, $M_{\rm GT,sd}$, $M_{\rm T,sd}$ and $\mathcal{M}^{0N}$ for $^{76}$Ge evaluated with different regulator cutoffs in the $0\nu\beta\beta$-decay operator. The black arrow indicates the effect of the SRG-evolution of the operator, so the bare operators are indicated by green lines and the SRG-evolved operators by red lines. We show the NMEs computed at $e_\text{max}=12$ (solid lines) and $e_\text{max}=6$ (dotted lines). We also include the dipole parameterization scheme, indicated by light and dark blue bars. The matrix element $M_{\rm F,sd}$ includes the factor $-(g_{\rm V}/g_{\rm A})^2$.}
    \label{fig:cutoff_plot}
\end{figure}

\section{SRG/Regulator analysis}
In this section, we give a more detailed analysis of the effect of the different regularization schemes for the short-range $0\nu\beta\beta$-decay operators as well as their SRG-evolution when using the N$^3$LO$_\text{LNL}$ interaction. In Fig. \ref{fig:cutoff_plot}, we show the values of $M_{\rm F,sd}$, $M_{\rm GT,sd}$, $M_{\rm T,sd}$ and $\mathcal{M}^{0N}$ in the isotope $^{76}$Ge for different cutoffs in the local regulator function for the transition operators, together with the effect of the SRG evolution of the operators indicated by the black arrow. For comparison, we also show the NMEs computed using the dipole parameterization. We see that the NMEs change significantly with the cutoff, although the dependence is slightly reduced when the operators are SRG-evolved. 

In Fig.~\ref{fig:cutoff_plot}, the NME exhibits its largest cutoff dependence at $550$ MeV, where a local minimum appears for the F, GT, and T operators. In contrast, the results between 450–500 MeV and 650–700 MeV are comparatively stable. This behavior cannot be attributed to the truncation of the model space, as the location of the minimum does not shift with $e_\text{max}$.

Imposing a maximum single-particle energy $e_\text{max}$ also imposes an approximate maximum single-particle momentum $p_\text{max}$, which acts as a regulator. For $\hbar\omega=15 $ MeV and $e_\text{max}=12$, we find $p_\text{max}\sim620$ MeV, potentially competing with the regulator of the transition operator when using cutoffs in the 550–650 MeV range, whereas the 450–500 MeV and 650–700 MeV regions lie safely below and above $p_\text{max}$, respectively. At $e_\text{max}=6$, with $\hbar\omega=15$ unchanged, $p_\text{max}\sim460$ MeV; if the minimum at 550 MeV were caused by the $e_\text{max}$ truncation, its position would likely shift leftward in Fig.~\ref{fig:cutoff_plot}. Since no such shift occurs, the truncation of the model space cannot explain the strong cutoff dependence near the value 550 MeV. Further investigation into the origin of this behavior is warranted, for example by analyzing the distribution of the NMEs in momentum space or by examining the role of induced contributions from the SRG flow to the transition operators, either through variation of the SRG scale or the inclusion of induced three-body terms via normal ordering.

\section{Sterile Neutrino Constraints}

{

In this section, we provide further details on the methodology used to construct constraints on $|U_{e4}|^2$ as function of the sterile neutrino mass assuming a saturating contribution to the $0\nu\beta\beta$-decay rate.

A Bayesian methodology is applied, using the framework laid out in Ref.~\cite{Biller2021,Shickele2026} to combine results from several recently completed or ongoing $0\nu\beta\beta$-decay experiments into the global limit. The likelihood function for LEGEND-200 ($^{76}$Ge), is taken from the posterior obtained in Ref.~\cite{LEGEND2025}, which also includes results from MAJORANA~\cite{MAJORANA2023} and GERDA~\cite{GERDA2020} in its analysis to obtain a half-life limit of $T_{1/2}^{0\nu} > 1.9 \times 10^{26}$ yr. Similarly for CUORE~\cite{CUORE2024} ($^{130}$Te) and KamLAND-Zen~\cite{KLZ2025} ($^{136}$Xe), we utilize the posterior and $\Delta\chi^2$ profile obtained by the collaborations as the experimental likelihoods, leading to limits of $T_{1/2} > 3.8 \times 10^{25}$ yr and $T_{1/2} > 3.0 \times 10^{26}$ yr respectively. The EXO-200~\cite{Exo2019} ($^{136}$Xe) likelihood function is built based on a simple Poisson counting analysis, marginalizing over the background uncertainties, for both Phase I and II. This leads to a somewhat stronger limit of $T_{1/2}^{0\nu} > 4.3 \times 10^{25}$ yr compared to that of $T_{1/2}^{0\nu} > 3.5 \times 10^{25}$ yr found by the EXO-200 collaboration, likely due to the difference in analysis procedure employed here. However, this discrepancy has a negligible impact on the final combined constraints.

Combining the experimental likelihoods obtained above, through their product,
\begin{align}
    \like_{\rm Combined} = \prod_{\rm Exps} \like_{\rm Exp},
\end{align}
leads to our final combined likelihood. This is further combined with a $\Gamma^{0\nu}$ uniform prior, consistent with previous prior choices in the literature~\cite{CUPIDMo2022,CUORE2024,GERDA2020,Biller2021,KLZ2025,Zhang2016,LEGEND2025}, leading to our final global $0\nu\beta\beta$-decay Bayesian constraints on the $|U_{e4}|^2$\,-\,$m_4$ parameter space in Fig.~\ref{fig:heavy-neutrino-limits}. Further details regarding the limit-setting procedure and experimental likelihoods can be found in Ref.~\cite{Shickele2026}.

\end{document}